\magnification\magstep1
\font\BBig=cmr10 scaled\magstep2


\def\title{
{\bf\BBig
\centerline{
The symmetries}
\bigskip
\centerline{of the}
\bigskip
\centerline{Manton superconductivity model
}
}} 


\def\authors{
\centerline{
M. HASSA\"INE\foot{e-mail: hassaine@univ-tours.fr},
P.~A.~HORV\'ATHY\foot{e-mail: horvathy@univ-tours.fr}}
\bigskip
\centerline{
D\'epartement de Math\'ematiques}
\medskip
\centerline{Universit\'e de Tours}
\medskip
\centerline{Parc de Grandmont,
F--37200 TOURS (France)
}
}

\def\runningauthors{
Hassa\"\i ne \& Horv\'athy
}

\def\runningtitle{
The symmetries of
the Manton model
}


\voffset = 1cm 
\baselineskip = 14pt 
\headline ={
\ifnum\pageno=1\hfill
\else\ifodd\pageno\hfil\tenit\runningtitle\hfil\tenrm\folio
\else\tenrm\folio\hfil\tenit\runningauthors\hfil
\fi
\fi}

\nopagenumbers
\footline={\hfil} 


\def\and{\qquad\hbox{and}\qquad}
\def\where{\qquad\hbox{where}\qquad}

\def\kikezd{\parag\underbar} 

\def\IR{{\bf R}}
\def\IZ{{\bf Z}}
\def\smallover#1/#2{\hbox{$\textstyle{#1\over#2}$}}
\def\2{{\smallover 1/2}}
\def\ccr{\cr\noalign{\medskip}} 
\def\parag{\hfil\break} 
\def\={\!=\!}
\def\p{\partial}

\def\D{{\cal D}}

\def\const{{\rm const}}
\def\eqletter#1{
\leqno(\the\ch.\the\eq{#1})
}


\newcount\ch 
\newcount\eq 
\newcount\foo 
\newcount\ref 

\def\chapter#1{
\parag\eq = 1\advance\ch by 1{\bf\the\ch.\enskip#1}
}

\def\equation{
\leqno(\the\ch.
\the\eq)\global\advance\eq by 1
}

\def\foot#1{
\footnote{($^{\the\foo}$)}{#1}\advance\foo by 1
} 

\def\reference{
\parag [\number\ref]\ \advance\ref by 1
}

\def\eqletter#1{
\leqno(\the\eq{#1})
}

\ch = 0 
\foo = 1 
\ref = 1 


\title
\vskip10mm
\authors
\vskip.20in

\parag{\bf Abstract.}
{\it The symmetries and
conserved quantities of Manton's modified superconductivity
model with  non-relativistic Maxwell-Chern-Simons dynamics
(also related to the Quantized Hall Effect)
are obtained in the ``Kaluza-Klein type'' framework of Duval et al.}

\vskip15mm
\noindent
(\the\day/\the\month/\the\year)
\bigskip
\medskip\noindent
\bigskip\noindent
\vskip35mm

\noindent
Key words : Landau-Ginzburg theory,
nonrelativistic symmetries, Kaluza-Klein theory.
\vskip5mm
\vfill\eject

\chapter{Introduction}

 Recently [1], Manton proposed a modified version
of the Landau-Ginzburg theory of superconductivity.
His equations,
defined on $(2+1)$-dimensional non-relativistic
spacetime parametrized by $\vec{x}$ and $t$, read
$$
i\gamma \D_t\Phi=
-{1\over2}\overrightarrow{\D}^2\Phi
-{\lambda\over4}\big(1-|\Phi|^2\big)\Phi,\hfill
\qquad\hbox{NLS}
\equation
$$
\null\vskip
-10mm
$$
\epsilon_{ij}\partial_{j}{\cal B}=
{\cal J}_{i}-J^{T}_{i}+2\kappa\,\epsilon_{ij}\,{\cal E}_j,
\qquad\hbox{Amp\`ere--Hall law}
\equation
$$
\null\vskip
-10mm
$$
2\kappa{\cal B}=\gamma\big(1-|\Phi|^2\big).
\qquad\hbox{Gauss' Law}
\equation
$$

Here $\gamma>0$, $\lambda>0$ and $\kappa\in\IR$ are constants,
${\cal B}=\epsilon^{ij}\partial_i{\cal A}^j$
and
$\overrightarrow{\cal E}=
\vec\nabla{\cal A}_t-\partial_t\overrightarrow{\cal A}$
are the ``statistical''  magnetic and the electric fields,
respectively, associated with the
vector potential $({\cal A}_t,\overrightarrow{\cal A})$.
The covariant derivatives mean
$\D_\alpha\Phi=\partial_\alpha\Phi-i{\cal A}_\alpha\Phi$;
the current is
$$
{\cal J}_{\alpha}={1\over2i}\big[\Phi^\star{\cal D}_{\alpha}\Phi
-\Phi({\cal D}_{\alpha}\Phi)^\star\big],
\equation
$$
$\alpha=t, i$. The
new ingredient is the transport current
$\vec{J}^{T}$ (a constant vector).

The matter field $\Phi$
satisfies hence a non-linear Schr\"odinger equation,
as in the
non-relativistic Chern-Simons theory of Jackiw and Pi [2].
The Maxwell term familiar from
the static Landau-Ginzburg theory
only enters Amp\`ere's law, (1.2), and
is  missing from Gauss' law, (1.3).
In the absence of a magnetic field  and of a transport
current, Eq. (1.2) reduces to the
 off-diagonal relation between the current and
 the electric field,
$$
{\cal J}_{i}=-2\kappa\,\epsilon_{ij}\,{\cal E}_j,
\leqno{(1.2')}
$$
which is Hall's law. The Manton model is in fact closely related to
the ``Landau--Ginzburg'' theory of the Quantized Hall Effect (QHE)
 [3], [4], and has indeed been used in this context [5].

The  form of the system (1.1-3)
is dictated by the requirement of galilean
covariance [1], [6];
it examplifies a  Galilei-invariant electromagnetic theory
of the magnetic type [7].

 To make the magnetic field vanish at infinity,
the particle density, $\varrho=\vert\Phi\vert^2$, has to tend
to $1$ rather than to zero when $r\to\infty$ by Eq. (1.3).
Similarly, it follows from Eq. (1.2)
that $\vec{\cal J}\to \vec{J}^T$ at infinity.

The system (1.1-3)
admits a surprising six-parameter algebra of symmetries [6], [8].
The first three are ordinary space and time translations.
The three further symmetries (related to those
 a constant external
electromagnetic field [9] and called
``hidden boosts and rotations'') are more subtle, see Section 5.

\goodbreak
The associated conserved quantities were obtained
in [6] and [8]. The procedure is somewhat tricky in that the
naive energy momentum tensor is not gauge-invariant and the associated
integrals do not converge. It has therefore to be ``improved''.
It is natural to inquire about the possibility
of obtaining
these improved expressions from first principles.

Another surprise is that the momenta
 satisfy the anomalous commutation relation [10]
$$
\big\{{p}_{1}, {p}_{2}\big\}=
\gamma\int\!Bd^2x
=2\pi\, n\,\gamma\,
\qquad
n\in\IZ,
\equation
$$
rather then commute, as ordinary translations do.
This relation is very important, since it can been used to
explain the quantization of the Hall conductivity in the QHE
[4].

In this paper, we explain these results
 using the ``non-relativistic Kaluza-Klein-type''
framework of Duval et al. [11].
The  well-known relativistic case is only
recalled  to motivate our arguments; proofs are only provided in the
non-relativistic context, which is our main concern here.

In the  approach of Ref. [11],
the $(2+1)$-dimensional dynamics is lifted
 to a 4-dimensional Lorentz manifold $(M, g)$ carrying a
covariantly constant lightlike vector $\xi$, referred to as
the ``Bargmann space''. Physics in ``ordinary'' space is recovered
by reduction along  $\xi$.
Owing to the singular character of the projection, our
systems defined on Bargmann space can only partially
be derived from an action principle,
forcing us to work mostly with the equations of motion.

In the simplest case (Case A), the Bargmann space is
Minkowski space, and we get a variant of the Jackiw-Pi theory [2],
[12].
In Case B, the external fields get included into the metric;
the reduction yields the Chern-Simons theory  in external fields [9], [13].
Our clue is that the transport terms behave precisely as
external electric and magnetic fields, so that
the metric B also describes the Manton model~!
Then our THEOREM1 states
that any $\xi$-preserving isometry of Bargmann space is a symmetry
for the reduced system.

In case A, the $\xi$-preserving isometries (resp. conformal transformations)
of $M$ form the extended Galilei (resp. Schr\"odinger [14]) group
[11].
For metric B, the conformal transformations form
 the ``hidden Schr\"odinger algebra'' [9], [13].
Describing the symmetries of the Manton model requires hence
 selecting the isometries of metric B. These form
 a {\it seven} parameter group, namely those
found in Ref. [6], augmented with the vertical translations.

Thus, the ``hidden''
symmetries are  also ``geometric'', but with respect to
another geometry.

Interestingly, the problem of
lifting the symmetries from ordinary space-time to
Barg\-mann space amounts to studying the symmetries
in a fixed background field, as discussed by Forg\'acs and Manton,
and by Jackiw and Manton [15].

The Bargmann framework also
allows us to derive a symmetric, conserved
energy momen\-tum tensor.
Then the geometric version of N{\oe}ther's theorem
(THEOREM3) associates a conserved quantity to each Killing vector
of Bargmann space, yielding, without further ``improvement'',
 the same conserved quantities as found before in Refs. [6] [8].
These facts underline the importance
of finding the ``good'' lift of the space-time transformations.

Our notation are as follows.
On ordinary space-time $Q$~: label $\alpha, \beta=t, i$.
Vectors $X=(X^\alpha)$;
the generators of the Galilei group~: upper-case letters, e. g.
$\overrightarrow{P}=(P^i)$, $\overrightarrow{G}$, $i=1, 2$ etc.;
generators of the hidden Galilei group~:
upper-case calligraphic letters,
e. g., $\overrightarrow{\cal P}$,
$\overrightarrow{\cal G}$.
Fields~: upper-case letters, e. g.
$A_{\alpha}$, $F_{\alpha\beta}$.
Conserved quantities~: lower-case letters, e. g.
$n$, $h$, $p_{i}$, $i=1, 2$, etc.
A general Lorentz 4-manifold: $(M, g_{\mu\nu})$.
On a Bargmann space
$(\widehat{M}$, $\hat{g}_{\mu\nu}, \xi)$
with special metric (2.2) below~:
``hat'' and label $\mu, \nu = t, i, s$.
Vectors $\hat{X}=(\hat{X}^\mu)$.
E. g., lift of an ordinary translation~:
$\widehat{\overrightarrow{P}}=(\widehat{P}^\mu)$;
lift of a hidden translation
$\widehat{\overrightarrow{\cal P}}$.
Fields~: lower-case letters; e. g. $a_{\mu}$, $f_{\mu\nu}$, etc.
On Minkowski space $\widetilde{M}$, $\tilde{g}_{\mu\nu}$~: ``tilde''
and label $\mu, \nu$.
E. g. lift of an ordinary translation
 $\widetilde{\overrightarrow{P}}$.

\goodbreak
\chapter{``A Kaluza-Klein'' framework for Maxwell-Chern-Simons theory}

\kikezd{I. General theory}.
 In relativistic Kaluza-Klein theory [16], electromagnetism
is described by a
Lorentz manifold ${M}$;
ordinary [relativistic] spacetime is the quotient
of $M$ by a {\it spacelike} fibration. To get electromagnetism
in the plane,
we chose $M$ to be $\IR^4$ with coordinates $x^\mu$
($\mu=\alpha, {5}$, $\alpha=0, 1, 2$)
and the metric
$$
\big(\widetilde{g}_{\alpha\beta}+A_{\alpha}^{ext}
A_{\beta}^{ext}\big)dx^\alpha dx^\beta
+
dx^{5}\big(A_{\alpha}^{ext}dx^\alpha+dx^{5}\big),
\equation
$$
where $\widetilde{g}_{\alpha\beta}$ is the (Minkowski) metric on
$(2+1)$ dimensional ordinary space-time $Q$;
 $A_{\alpha}$ represents
the electromagnetic vector potential. The space-like direction
to be factored out is generated by
the Killing vector $\xi=\p_{5}$.

Gauge theory admits another geometric description, namely using the
language of fiber-bundles
[17],[18], [DH]. The external electromagnetic field is represented by a
connection $1$-form $\varpi$ on a principal $\IR$ (or U$(1)$) bundle
$M$ over spacetime $Q$, whose curvature is the electromagnetic two-form,
$d\varpi=F$. The vector potential $A_{\alpha}$ is the pull-back of
the connection form $\varpi$ by a section of the bundle.
This approach makes no reference to any metric, and is therefore valid
in both the relativistic and the nonrelativistic context.

A  Kaluza--Klein type framework
for non-relativistic physics  was given in [11].
Let us consider a 4-manifold $M$, which is
endowed with a Lorentz metric
of signature $(-, +, +, +)$ and also carries a covariantly constant null
vector, $\xi=(\xi^\mu)$. The quotient
of $M$ by the flow of $\xi$,
 denoted by $Q$, is a $(2+1)$-dimensional manifold with
a Newton-Cartan structure, i.e., a non-relativistic spacetime [11].
As found long time ago [20], the most general ``Bargmann''
4-space has the form
$$
g_{ij}dx^idx^j+2dt\big[ds+(1/\gamma)A^{ext}_{i}dx^i\big]
+2(1/\gamma)A^{ext}_{t}dt^2.
$$

Here the ``transverse metric''
$g_{ij}$, as well as the ``vector'' and ``scalar'' potentials
$A^{ext}_{i}$ and  $A^{ext}_{t}$,
are functions of ``galilean time'' and ``position'',
 $t$ and $\vec{x}$.
 $\xi=\p_{s}$ is  a covariantly constant null vector.

In this paper, we only consider Brinkmann  metrics with
flat transverse space,
$$
\hat{g}_{\mu\nu}dx^\mu dx^\nu
=
\delta_{ij}dx^idx^j+2dt\big[ds+(1/\gamma)A^{ext}_{i}dx^i\big]
+2(1/\gamma)A^{ext}_{t}dt^2.
\equation
$$
All such metrics
can be viewed as defined on the
same manifold (topologically $\IR^4$), obtained
by distorting the ``vertical'' components of
the Minkowski-space metric.
$$\eqalign{
&\hat{g}_{\mu\nu}=\tilde{g}_{\mu\nu}+\eta_{\mu\nu},
\cr
&\tilde{g}_{\mu\nu}dx^\mu dx^\nu
=d{\vec{x}\strut}^{2}+2dtds,
\qquad
\eta_{\mu\nu}dx^\mu dx^\nu=2A^{ext}_{\alpha}dx^\alpha dt.
\cr}
\equation
$$
The fields
$$
\vec{E}^{ext}
=
-\partial_t\vec{A}^{ext}+\vec{\nabla}A^{ext}_{t}
\and
B^{ext}=\vec{\nabla}\times\vec{A}^{ext}
\equation
$$
have been interpreted as external electric and magnetic fields,
respectively [11].

In the relativistic case,  the fiber bundle approach
can be recovered from that of  Kaluza-Klein~: the total space,
$M$, is itself a fiber bundle
with typical fiber generated by $\xi=\p_{5}$.
$$
\varpi=i_{\xi}g\equiv g(\xi,\,\cdot\,)
\equation
$$
is a connection form on this bundle. Contracting (2.1) with
$\xi=\p_{5}$ yields indeed the standard expression
$\varpi=A_{\alpha}^{ext}dx^\alpha+dx^{5}$.

In the non-relativistic case
the formula (2.5)  breaks down,
because the vertical fibration is  lighlike~:
$\varpi(\xi)=\xi_{\mu}\xi^\mu=0$, contradicting the condition
$\varpi(\xi)=1$ required for a connection form [17].
Put another way, the tangent space to the bundle
 can not be decomposed into the direct sum of a horizontal
subspace and the vertical subspace since $\xi$ is itself horizontal.

\kikezd{II. Non-relativistic Chern-Simons theory}.
 Let us now present our non-relativistic
Maxwell-Chern-Simons field theory on $M$.
Let $f=\2f_{\mu\nu}dx^\mu\wedge dx^\nu$
be a closed 2-form. and $j=\big(j^\sigma\big)$ a vector on $M$.
(Locally $f_{\mu\nu}=2\partial_{[\mu}a_{\nu]}$).
Slightly generalizing the procedure proposed in Ref. [12],
we

(I) posit the
generalized Maxwell-Chern-Simons Field-Current Identities
(FCI) on Barg\-mann space
$$
\sqrt{-g}\,\epsilon_{\mu\nu\rho\sigma}\xi^\rho
\nabla_{\omega}f^{\omega\sigma}
+
2\kappa f_{\mu\nu}
=
-\sqrt{-g}\,\epsilon_{\mu\nu\rho\sigma}\xi^\rho
j^\sigma,
\equation
$$
where $\nabla_{\mu}$ is the covariant derivative w. r. t. the
metric $\hat{g}_{\mu\nu}$;

(II) consider the non-linear wave equation
for a scalar field $\phi$ on $M$,
$$
\D_{\mu}\D^{\mu}\phi-{R\over6}\phi
-
2{\delta U\over \delta\phi^\star}=0,
\equation
$$
\vskip-2mm\noindent
where $\D_{\mu}$ is the metric and gauge covariant derivative,
\vskip-5mm
$$
\D_{\mu}=\nabla_{\mu}-ia_{\mu},
\equation
$$
$R$ is the scalar curvature of the Bargmann space $M$,
and $U=U(\vert\phi\vert^2)$ is some scalar potential;

(III) couple
Eqns. (2.8) and (2.9) according to
$$
j^\mu={1\over2i}\big[\phi^\star\D^\mu\phi-\phi(\D^\mu\phi)^\star\big].
\equation
$$

Requiring $\phi$ to be equivariant,
$$
\xi^\mu{\cal D}_{\mu}\phi=i\gamma\,\phi,
\equation
$$
this system of equations will project into one on $Q$.
Indeed, contracting Eq. (2.8) with the vertical vector
$\xi$, the antisymmetry implies that
$
f_{\mu\nu}\xi^\nu=0.
$
But $f_{\mu\nu}$ also satisfies, by construction, the
homogeneous Maxwell equations
$
\partial_{[\rho}f_{\mu\nu]}=0.
$
Therefore, the Lie derivative of the two-form $f$ 
by $\xi$ vanishes,
$
L_\xi f=0.
$
It follows that the field strength
$f$
is the lift to $M$ of a two-form
$F=\2F_{\alpha\beta}dx^\alpha\wedge dx^\beta$ on $Q$.
Similarly, using that $\xi$ is covariantly constant, it follows from Eq. (2.8)
 that $L_\xi j=0$.
The current $j^\mu$ projects therefore to one on
$Q$ we denote by
$J=(J^\alpha)$.
Eq. (2.8) descends therefore to $Q$, providing us with
Maxwell-Chern-Simons equations in $(2+1)$ dimensions.

Finally, owing again to equivariance and the form of the metric,
the non-linear wave equation (2.9) projects to
one on $Q$.

For simplicity, we  only consider the
symmetry-breaking fourth-order potential
$$
U(\vert\phi\vert^2)={\lambda\over8}\big(1-\vert\phi\vert^2\big)^2.
\equation
$$

\goodbreak
\kikezd{III. Examples}.
Let us now consider some examples.
\kikezd{Case A}.
 The simplest choice is Minkowski space,  $M=\widetilde{M}$,
$$
\widetilde{g}_{\mu\nu}dx^\mu dx^\nu=(d\vec{x})^2+2dtds,
\qquad\hbox{(Minkowski space)}
\equation
$$
Then, setting
$
\Phi=e^{-i\gamma s}\phi,
$
our Eqns. (2.8-9) reduce to those of Jackiw and Pi
[2] with an additional magnetic Maxwell term and a different
potential,
$$
\left\{\eqalign{
&B=-{\gamma\over2\kappa}\,\varrho,
\ccr
&\epsilon_{ij}\partial_{j}B=
J_{i}+2\kappa\,\epsilon_{ij}\,E_j,
\ccr
&i\gamma D_t\Phi=\left[
-{1\over2}\,\vec{D}^2
-{\lambda\over4}\big(1-\vert\Phi\vert^2\big)\right]\Phi,}\right.
\equation
$$
where
$
\varrho=\Phi^*\Phi
$
and
$
\vec{J}={1\over 2i}\left[
\Phi^*\vec{D}\Phi-\Phi(\vec{D}\Phi)^*\right]$
and $D_{\alpha}=\p_{\alpha}-iA_{\alpha}$.

\goodbreak
\kikezd{Case B}.
Let us now consider the special Brinkmann
metric $\hat{g}_{\mu\nu}$ in (2.2) on $\IR^{4}$,
with
$$\left\{\matrix{
&A^{ext}_{i}=\2\epsilon_{ij}x^jB^{ext},\qquad
\hfill
&B^{ext}=\const,\hfill
\ccr
&A^{ext}_{t}=\vec{x}\cdot{\vec{E}}^{ext},
\hfill
&{\vec{E}}^{ext}=\const.\hfill
\cr}\right.
\equation
$$
Such a metric has vanishing scalar curvature, $R=0$.
Since the only non-vanishing components of the inverse metric
are
$\hat{g}^{ij},\, \hat{g}^{is}=-A^{ext}_i,\,
\hat{g}^{ss}=-2A^{ext}_{t}-(A^{ext})^i$
and $\hat{g}^{ts}=1$, we find that
the extra components of the metric simply modify,
after reduction, the covariant derivative as
$$
{\cal D}_{\alpha}\equiv\p_{\alpha}-iA_{\alpha}-iA^{ext}_{\alpha}=
D_{\alpha}-iA^{ext}_{\alpha},
\equation
$$
$\alpha=t,\, i$. Our equations become hence
$$\left\{\eqalign{
&B=-{\gamma\over2\kappa}\,\varrho,
\ccr
&\epsilon_{ij}\partial_{j}B=
{\cal J}_{i}+2\kappa\,\epsilon_{ij}\,E_j,
\ccr
&i\gamma\underbrace{\big(\p_{t}-iA_{t}-iA^{ext}_{t}
\big)}_{{\cal D}_{t}}\Phi
=
\left[-{1\over2}\big(
\underbrace{\vec\nabla-i\vec{A}-i\vec{A}^{ext}}_{\overrightarrow{\cal D}}
\big)^{2}\Phi
-{\lambda\over4}\big(1-\vert\Phi\vert^2\big)\right]\Phi.
\cr}\right.
\equation
$$

The fields $B$ and $\vec{E}$
 here only involve the ``statistical gauge field''
$A_{\alpha}$
but not the background terms:
$B=\epsilon_{ij}\partial_iA_{j}$,
$\vec{E}=\vec\nabla A_t-\partial_t\vec{A}$.
The external field enters
the non-linear Schr\"odinger equation, though,
and also change the current (1.4),
$
{\cal J}_{\alpha}=J_{\alpha}-A^{ext}_{\alpha}\vert\Phi\vert^2.
$
Consistently with the interpretation of $\vec{E}^{ext}$
and ${B^{ext}}$ in [11],
these equations describe non-relativistic
Chern-Simons vortices in a constant external electric
and magnetic field [9] (again with an additional magnetic
Maxwell term).

This same system admits also another interpretation.
For
$$
A^{ext}_{t}={1\over2\kappa}\vec{x}\times\vec{J}^T,
\qquad
A^{ext}_{i}=-{\gamma\over4\kappa}\,\epsilon_{ij}x^j,
\equation
$$
we get
$$
{B^{ext}}=
{\gamma\over2\kappa},
\qquad
E^{ext}_{i}=-{\epsilon_{ij}{J^{T}}_{j}\over2\kappa}.
\equation
$$
 Setting
 ${\cal A}_{\alpha}=A_{\alpha}+A^{ext}_{\alpha}$,
 we have
$$
{\cal B}=B+{\gamma\over2\kappa}
\and
{\cal E}_{i}=E_{i}-\epsilon_{ij}J^T_{j},
\equation
$$
so that in terms of the curly quantities
the eqns. (2.18) become those of Manton, (1.1-3)~!
\goodbreak

\goodbreak
\chapter{Variational aspects}

It is natural to ask whether the posited field equations
 come from a variational principle.
A system similar to ours was  considered by Carroll, Field and Jackiw
in Ref. [21]. Adapting their approach to our case, let us start with
a with Lorentz $4$-manifold  $M$,
endowed with a covariantly constant
vector $\xi^\mu$, and add
 tentatively a Chern-Simons-type
term to the usual matter -- Maxwell
Lagrangian, $L=L_{1}+L_{2}$, where
$$\eqalign{
L_{1}&={1\over4}f_{\mu\nu}f^{\mu\nu}
+\2(\D_\mu\phi)^*\,\D^\mu\phi
+{R\over12}\,|\phi|^2
+U(|\phi|^2),
\ccr
L_{2}&={\kappa\over2}\,
\epsilon^{\mu\nu\rho\sigma}\xi_\mu a_\nu f_{\rho\sigma}.
\cr}
\equation
$$

The resulting field equations read
$$
\partial_\mu f^{\mu\nu}
+\kappa\,\sqrt{-g}\,\epsilon^{\mu\rho\sigma\nu}\xi_\mu f_{\rho\sigma}
={\ \delta\over\delta a_\nu}\big(L_{1}\big)
=-j^\nu,
\equation
$$
supplemented with the matter equation (2.5).

In order to relate this theory to one in one less dimensions,
let us assume that $\phi$ is equivariant, (2.12),
and that $f_{\mu\nu}$ is the lift from $Q$
of a two-form  $F_{\alpha\beta}$
\foot{In the appoach presented in {\rm I}, this follows
automatically from the field equation.}.
Hence
$
f_{\mu\nu}\xi^\mu=0.
$
\goodbreak

The field equations (3.2)  are similar to those in (2.8),
except for the
``wrong'' position of the $\epsilon_{\mu\nu\rho\sigma}$ tensor.
To compare the two theories, let us transfer the
$\epsilon_{\mu\nu\rho\sigma}$ to the other side of Eq. (3.2)
and contract with $\xi^\rho$ to get, using
$
f_{\mu\nu}\xi^\mu=0,
$
$$
\sqrt{-g}\,\epsilon_{\mu\nu\rho\sigma}\,\xi^\rho\partial_\tau f^{\tau\sigma}+
\kappa\,(\xi_\rho\xi^\rho)f_{\mu\nu}
=\sqrt{-g}\,\epsilon_{\mu\nu\rho\sigma}\,\xi^\rho j^\sigma.
\equation
$$

If $\xi$ is {\it spacelike} (or {\it timelike}),
 it can be normalised as
$\xi_\mu\xi^\mu=\pm1$. Then Eq. (3.3) is (possibly up to a sign)
our equation (2.8).
In the spacelike case, for example,
we get a well-behaved {\it relativistic}
model~: the quotient
is a Lorentz (or a euclidean)  manifold.
Let $M$ be, for example,  Minkowski space with metric
$dx^2+dy^2-dz^2+dw^2$. The vector $\xi=\partial_w$ is space-like
and covariantly constant.
The quotient is $(2+1)$-dimensional Minkowski space
 with metric $dx^2+dy^2-dz^2$ and
the equations (2.8)-(2.9) project to
$$
\left\{\eqalign{
&\partial_\alpha F^{\alpha\gamma}
+\kappa\,\epsilon^{\alpha\beta\gamma}F_{\alpha\beta}
=J^\gamma,
\cr
&\D_{\alpha}\D^\alpha\Phi
-
2{\delta U\over\delta\Phi^\star}=0.
\cr}\right.
\equation
$$
These are indeed the correct Maxwell-Chern-Simons and Klein-Gordon equations
for a  relativistic model in (2+1)-dimensional flat space [22].

 If, however, $\xi$ is {\it lightlike},
$\xi_\mu\xi^\mu=0$,
then the $f_{\mu\nu}$ has a {\it vanishing} coefficient,
and Eqn. (3.3) does not reproduce
those in (1.2)-(1.3).

In conclusion, (3.1) is
a correct Lagrangian in the relativistic case but
fails to work in the lightlike case, which is precisely
our case of interest here.
It is worth remarking, however, that the non-linear wave equation,
(2.9), {\it is} correctly
reproduced by variation of the ``partial action''
$$
S=\int_M\left\{
\smallover1/{4}f_{\mu\nu}f^{\mu\nu}
+\2(\D_\mu\phi)^*\,\D^\mu\phi
-\2j^T_{\mu}{j^T}^\mu
+{R\over12}\,|\phi|^2
+U(|\phi|^2)
\right\}\sqrt{-g}\,d^4\!x.
\equation
$$
In order to make the integral converge,
 we have added here the (constant) transport term  to the Lagrange
 density,
$$-\2j^T_{\mu}{j^T}^\mu,
\where
(j^T)^{t}=\gamma
\quad
(j^T)^{i}=(J^T)^{i},
\qquad
(j^T)^{s}=0.
$$
In fact,
$(\D_\mu\phi)^*\,\D^\mu\phi\to j^T_{\mu}{j^T}^\mu$
as $r\to \infty$.
We also included a Maxwell term for future convenience,
see Section 6.

\goodbreak
\chapter{Space-time symmetries}

\kikezd{I. Symmetries in ordinary space}.
Let us now discuss the space-time symmetries.
 In the Forg\'acs--Manton--Jackiw approach [15],
(infinitesimal) symmetries are represented by
vector fields $X=(X^\alpha)$ on space-time.
In the relativistic context considered by
the above authors, these are typically
Killing vectors of the space-time metric, which
leave the kinetic term invariant and hence act as symmetries for a
free system.
In the presence of an external electromagnetic field,
however, only those vector fields remain symmetries
for which the change of the external vector potential
$A^{ext}_{\alpha}$
can be compensated by a suitable gauge transformation,
$$
L_{X}A^{ext}_{\alpha}=\p_{\alpha}W
\qquad
\alpha= i,\ t,
\equation
$$
for some compensating function $W(\vec{x}, t)$.
(Owing to the gauge freedom,
strict invariance,
$
L_{X}A^{ext}_{\alpha}=0,
$
would  be too restrictive.)
Using the identity
$$
\big(L_{X}A)_{\alpha}=
\p_{\alpha}\big(A_{\beta}X^\beta\big)+X^\beta F_{\beta\alpha}
$$
valid for any $1$-form,
this relation is readily seen to be equivalent to
$$
F^{ext}_{\alpha\beta}X^\beta=\p_{\alpha}\Upsilon
\where
\Upsilon=A^{ext}_{\alpha}X^\alpha-W.
\equation
$$

Note that while $A_{\alpha}$ and $W$ are gauge dependent, $\Upsilon$
represents the gauge independent respons of the field
to a symmetry transformation [15]. The respons function
$\Upsilon$ also appears in the
``spin from isospin contribution''
 in the associated conservation law, see Section 6.

When  several symmetries are present in the theory, they only form a
closed algebra when, for any two symmetries
$X_{1}$ and $X_{2}$ and compensating functions
$W_{1}\equiv W_{X_{1}}$ and
$W_{2}\equiv W_{X_{2}}$, the additional relations
$$
L_{X_{1}}W_{2}-L_{X_{2}}W_{1}=
W_{[X_{1},X_{2}]}
\equation
$$
or equivalently
$$
F_{\alpha\beta}^{ext}X^\alpha_{1} X^\beta_{2}
=\Upsilon_{[X_{1},X_{2}]}
\leqno{(4.3')}
$$
are also satisfied. Expressed using the
transport terms, our conditions (4.1) and (4.3) require
$$\eqalign{
&\vec{X}\times\vec{J}^{T}
=
\p_t\Upsilon,
\cr
&\epsilon_{ij}\big(X^t{J^{T}}^j-X^j{J^{T}}^t\big)
=\p_{i}\Upsilon.
\cr}
\equation
$$

\kikezd{II. The bundle picture}.
 The ordinary-space approach of Forg\'acs et al.
 can be translated into  fiber-bundle language
[18], [19].
A symmetry of the external electromagnetic field
 is a vector field $\hat{X}=(\hat{X}^\mu)$
on the bundle which is invariant w. r. t. the action of the structure
group on the fibers and which also leaves the connection form
invariant,
$$
L_{\hat{X}}\varpi=0.
\equation
$$
In terms of a local section $s:Q\to M$, this condition means precisely
(4.1) where $A^{ext}=s^\star\varpi$. The lift
admit the gauge-invariant expression
$$
\hat{X}=\overline{X}-\Upsilon^*
\equation
$$
where $\overline{X}$ is the horizontal lift of $X$,
$\varpi(\overline{X})=0$, and $\Upsilon^*$ denotes the fundamental
vectorfield [17] associated to $\Upsilon$. This
latter can be recovered from the lift according to
$$
\Upsilon=-\varpi(\hat{X}),
\equation
$$
since $\varpi(\Upsilon^\star)=\Upsilon$.
The consistency condition (4.3) means that the lifts close
into a Lie algebra which acts on $\hat{M}$,
$$
\big[\widehat{X}_{1}, \widehat{X}_{2}\big]
=
\widehat{\big[{X}_{1}, {X}_{2}\big]}.
\equation
$$

Eqn. (4.3') provides in fact a cohomological obstruction
for lifting the Lie algebra isomorphically from the base to
the bundle [23], [24].

\kikezd{III. The Kaluza-Klein approach}.
 In the relativistic case, the symmetry conditions
(4.1) or (4.2) are readily seen to be equivalent to requiring that
the lift $\hat{X}$ be an isometry of the Kaluza-Klein metric,
$$
L_{\hat{X}}g=0.
\equation
$$
In fact, $L_{\hat{X}}\varpi=L_{\hat{X}}\big(i_{\xi}g\big)
=i_{\xi}\big(L_{\hat{X}}g\big)$.
Then the gauge-invariant respons of the  field to a symmetry
transformation is recovered as
$$
\Upsilon=g_{\mu\nu}\xi^\mu\hat{X}^\nu
=\xi_{\nu}\hat{X}^\nu.
\equation
$$

Let us now turn to the  non-relativistic case.
The r\^ole of space-time
isometries is played here by Galilei transformations\foot{For
the purely quartic potential
$U=-(\lambda/8)\vert\phi\vert^4$,
the Chern-Simons system is symmetric with respect to
 ``non-relativistic conformal transformations'' [2], [13].}.
However,
since we are only interested in the  potential (2.13)
which manifestly breaks the conformal transformations,
we focus our attention to the isometries.
Thus, let $\hat{X}=(\hat{X}^\mu)$ be a Killing vector
of the Bargmann metric $\hat{g}_{\mu\nu}$
which also preserves the vertical vector $\xi$,
$$
L_{\hat{X}}\hat{g}=0,
\qquad
\big[\hat{X},\xi\big]=0.
\equation
$$

When $\xi$ is factored out, such a vector projects onto an
(infinitesimal) ``galilean isometry''  of space-time $Q$,
 we denote (with some abuse of notation) by
$X=(X^\alpha)$, ($\alpha=t, i$).
 In our case this simply means a
Galilei transformation of $(2+1)$--dimensional space-time.
(The general case is discussed in Ref. [11]).

Conversely, an infinitesimal Galilei transformation
$X^\alpha$ on flat space-time, $Q$
lifts, by construction, as the Killing vector
$\widetilde{X}^\mu$ on Minkowski space $\widetilde{M}$.
What about a more general Brinkmann metric (2.2)~?
Let us assume that  $X^\alpha$ lifts as a Killing vector
$\hat{X}^\mu$ to $(\hat{M}, \hat{g})$.
Since  $\widetilde{X}^\mu$ and
$\hat{X}^\mu$ are lifts to $\IR^4$ of the same Galilei
transformation,
$$
\hat{X}^\mu=\widetilde{X}^\mu+Y^\mu
\equation
$$
where $Y^\mu$ is vertical.
In our preferred local frame, we
denote the only non-vanishing component of $Y^\mu$ by $W$,
$Y^\mu=-W\xi^\mu$.
The Lie derivative of the Brinkmann metric (2.2) is hence
$$
\eqalign{
(L_{\widehat{X}}\hat{g})_{\mu\nu}
=
(L_{\widetilde{X}}\tilde{g})_{\mu\nu}
&+\widetilde{X}^\rho
\partial_\rho\eta_{\mu \nu}
+\eta_{\mu\rho}\partial_{\nu}\widetilde{X}^\rho+
\eta_{\rho\nu}\partial_{\mu}\widetilde{X}^\rho
\cr
&-\tilde{g}_{\mu s}\partial_{\nu}W-\tilde{g}_{s\nu}\partial_{\mu}W
-\eta_{\mu s}\partial_{\nu}W-\eta_{s\nu}\partial_{\mu}W.
\cr}
$$
Here $(L_{\widetilde{X}}\tilde{g})_{\mu\nu}=0$,
 since $\widetilde{X}$ is
Killing for Minkowski. The vanishing of
$(L_{\hat{X}}\hat{g})_{\mu\nu}$ requires thus
$$
\widetilde{X}^\rho\partial_\rho\eta_{\mu\nu}
-
\delta_{\mu o}\partial_{\nu}W-\delta_{\nu o}\partial_{\mu}W
+
\eta_{\mu\rho}\partial_{\nu}\widetilde{X}^\rho
+
\eta_{\rho\nu}\partial_{\mu}\widetilde{X}^\rho=0.
\equation
$$
This relation is  automatically satisfied
with the exception of the components
$(t,\alpha)$, for which it requires
$$
\widetilde{X}^\beta\p_{\beta}A^{ext}_{\alpha}
+
A^{ext}_{\beta}\p_{\alpha}\widetilde{X}^\beta
=
\p_{\alpha}W.
$$

On the l. h. s. here we recognize the Lie derivative
of $A^{ext}_{\alpha}$ w. r. t.
 the ``galilean isometry''
$X=(X^\alpha)$, which is hence a
symmetry for the  external electromagnetic field
in the sense of Forg\'acs-Manton-Jackiw [15],
as anticipated by the notation.

\goodbreak
\chapter{Symmetries of the field-theoretical system}

Now we prove the following

\kikezd{THEOREM 1}.
{\it Any $\xi$
preserving isometry of Bargmann space is a symmetry of our
coupled system of equations (2.8-9)}.
\vskip2mm

By a symmetry we mean here a transformation which
carries a solution into some other solution of the equations of motion.

Our theorem can be shown along the same lines as in Ref. [12].
Let us first consider the nonlinear wave equation (2.9).
It is easy to see that transforming the fields as
$$
\delta\phi=L_{\hat{X}}\phi,
\qquad
\delta a_{\mu}=L_{\hat{X}}a_{\mu},
$$
the new fields,
$$
\phi^*=\phi+\delta\phi
\and
a_{\mu}^*= a_{\mu}+\delta a_{\mu},
$$
are still solutions
of (2.9) for all isometries $\hat{X}^\mu$ of $(\hat{M}, \hat{g})$.

Next, the equivariance condition (2.9) plainly requires
$\hat{X}^\mu$ to be $\xi$-preserving. Then the current equation
(2.8) behave also correctly.

Finally, let us consider the Chern-Simons equations, (2.8).
Using
$$
\delta f_{\mu\nu}=L_{\hat{X}}f_{\mu\nu}
=f_{\mu\rho}\p_{\nu}\hat{X}^\rho
\hat{X}+
f_{\sigma\nu}\p_{\mu}\hat{X}^\sigma
+
\hat{X}^\rho\p_{\rho}f_{\mu\nu},
$$
Eq (2.8) becomes, for
$
f_{\mu\nu}^*=f_{\mu\nu}+\delta f_{\mu\nu},
$
$$
2\kappa
f_{\mu\nu}^*
=
-\sqrt{-g}\,\epsilon_{\mu\nu\rho\sigma}\xi^\rho\,
\big({j^\sigma}^*+\nabla_{\omega}{f^{\omega\nu}}^*
\big),
$$
i.e.,
$$
\sqrt{-g}\,\epsilon_{\mu\nu\rho\sigma}\xi^\rho\,
\nabla_{\omega}{f^{\omega\nu}}^*
+2\kappa f_{\mu\nu}^*
=
-\sqrt{-g}\,\epsilon_{\mu\nu\rho\sigma}\xi^\rho\,
{j^\sigma}^*,
$$
as required.

 To end the general theory, let us point out that
Bargmann-conformally related Barg\-mann manifolds
share the same symmetries.
This is explained by the geometric version [13] of the
``export-import'' procedure [9],
originally due to Niederer [25].
Let us indeed consider two Bargmann spaces
$(\hat{M}, \hat{g}, \hat{\xi})$
and
$(\widetilde{M}, \tilde{g},\xi)$\foot{Note that
 $\hat{\xi}=\tilde{\xi}=\p_{s}=\xi$ generates the
 vertical translations we denote also by $N$.}, and assume that they are
Bargmann-conformally related, i.e., there is a differentiable map
$\Psi~: \hat{M}\to \tilde{M}$
such that
$$
\Psi^*\tilde{g}=\Omega^2\hat{g}
\and
\Psi_{*}\tilde{\xi}=\hat{\xi},
\equation
$$
where $\Omega(t,\vec{x})$ is some positive function.
Then, the image by $\Psi$ of any $\xi$-preserving
conformal vectorfield $\hat{X}^\nu$ on $\hat{M}$,
$$
\big(\widetilde{\Psi_{*}X\big)}^\mu
=
{\p\tilde{x}^\mu\over\p\hat{x}^\nu}X^\nu,
\equation
$$
is a $\hat{\xi}$-preserving conformal vectorfield
on $\tilde{M}$. (The image of a
Killing vector for $\hat{g}$ may not be Killing for $\tilde{g}$, though).
The algebraic structure is preserved by the ``exportation'',
$$
\Psi_{*}[X_{(1)},X_{(2)}]=
\big[\Psi_{*}X_{(1)}, \Psi_{*}X_{(2)}\big].
\equation
$$

\goodbreak
\kikezd{Examples}.
Let us again consider our examples.

\goodbreak
\kikezd{Case A}.
The $\xi=\p_{s}$-preserving conformal vectors of Minkowski
space,
$$
\widetilde{X}^\mu
=
\pmatrix{-\chi t^2-\rho\,t-\epsilon\hfill\ccr
\Omega(\vec{x})
-\left(\2\rho+\chi t\right)\vec{x}
+t\vec{\beta}+\vec{\delta} \hfill \ccr
\2\chi\,|\vec{x}|^2-\vec{\beta}\cdot\vec{x}+\eta
 \hfill\cr
},
\equation
$$
where $\Omega\in{\rm so}(2),\,
\vec{\beta},\vec{\delta}\in\IR^2,\,
\epsilon,\chi,\rho,\eta\in\IR$,
interpreted as rotation ($\tilde{R}$), boost ($\tilde{G}$),
space translation ($\tilde{P}$),
time translation ($\tilde{H}$),
expansion ($\tilde{K}$), dilatation ($\tilde{D}$) and vertical translation
($N$).
Calculating the commutation relations
shows that this $9$-dimensional Lie algebra is indeed the
{\it centrally extended Schr\"odinger algebra}.
The central extension shows up in the
commutator of translations with boosts,
$$
\big[\underbrace{\hbox{translation}_{i}}_{\widetilde{P}_{i}},
\underbrace{\hbox{boost}_{j}}_{\widetilde{G}_{i}}\big]
=
-\delta_{ij}\big(\underbrace{\hbox{vertical translation}}_{N}\big).
\equation
$$

Projecting the algebra (5.4) into $Q$
(which amounts to keeping just the first two components), we get the
$8$-dimensional Schr\"odinger algebra of Ref. [14]~:
the central extension is lost under projection.

The $\xi$-preserving Minkowski-space isometries
form a $7$-dimensional algebra, namely
the planar {\it centrally extended Galilei algebra} (also called the
Bargmann algebra), consisting of
rotations, boosts, spatial and time translations as well as
``vertical'' translations (translations along $\xi$),
given by (5.4) with $\chi=\rho=0$.
Projecting the $\xi$-preserving Killing vectors into $Q$, we get
the planar Galilei algebra with $6$ generators,
whose commutation relations differ from those
on Bargmann space in that ordinary-space
boosts and translations  commute.

The potential (2.13)
breaks the conformal transformations to isometries.
Then Theorem1 implies that the extended Galilei algebra
is symmetry for the system (2.15).

\goodbreak
\kikezd{Case B}.
 The conformal vectors of  Metric B
form again a $9$-dimensional
Lie algebra, which is algebraically isomorphic to the Schr\"odinger algebra.
This  can either be shown by a lengthy
 direct calculation, or be derived by the
``export-import'' procedure [9] explained above. Consider
the mapping
$$
\Psi(t,\vec{x},s)=(T,\vec{X},S)
$$
presented in Eq. (7.1) below in the Appendix,
constructed of
(i) Niederer's transformation [25] which takes the free case
to an oscillator,
followed by (ii) a rotation which carries the oscillator
into a uniform magnetic field
and (iii) followed again by
 a boost which creates a non-zero
electric field.
Then $\Psi$ carries the Bargmann space
$\big(\widehat{M},\hat{g},{\xi}\big)$ of Case B
into  Minkowski space
$\big(\widetilde{M},\tilde{g},{\xi}\big)$
in a $\xi$--preserving manner.

The image by the inverse mapping $\Psi^{-1}$ of the generators (5.4)
 $\widetilde{X}^\mu$ of the
Schr\"odinger group of Case A is
a $9$-dimensional algebra defined on the same manifold $\IR^4$,
made of $\hat{\xi}$ (=$\xi$) preserving conformal vectors w. r. t.
the metric $\hat{g}_{\mu\nu}$ of Case B.
By (5.3), these generators satisfy by construction
the same commutation relations as their pre-images.
We call it therefore ``the hidden Schr\"odinger algebra''.
These formul{\ae} (presented in the Appendix)
 are rather complicated
nevertheless necessary to establish the
crucial relations (5.7) and (5.9) below. These latters provide
in turn the ``good'' lifts (5.8) and (5.10).

For the  quartic potential $-\lambda\Phi^4$, all  conformal
generators would act as symmetries [9] [13].
For the symmetry-breaking potential (2.13), only isometries
i. e.
solutions of the Killing equation
$$
L_{\hat{X}}\hat{g}=0
\equation
$$
qualify, though.
These are, first of all
``hidden translations'', $\widehat{\overrightarrow{\cal P}}$,
``hidden boosts'', $\widehat{\overrightarrow{\cal G}}$,
``hidden rotation'', $\widehat{{\cal R}}$
  and vertical translation
(listed in Eq. (7.2) in the Appendix).

Some of the generators can be replaced by more familiar expressions,
though.
A certain combination of ``hidden translations''
``hidden boost'' projects in fact to ordinary translations,
$$\hbox{$\matrix{
\underbrace{\hbox{(ordinary translation)}_{i}}_{\widehat{P}_{i}}
\ccr
=
\ccr
\underbrace{\hbox{(``hidden translation'')}_{i}}_{\widehat{\cal{P}}_{i}}
+{\gamma\over4\kappa}\,\epsilon_{ij}\,
\underbrace{\hbox{(``hidden boost'')}_{j}}_{\widehat{\cal G}_{j}},
\cr}
$}
\equation
$$
showing that we could have traded either the hidden translations
or the hidden boosts for ordinary translations and vice versa.
The lift to
 $(\hat{M}, \hat{g}_{\mu\nu})$  of ordinary translations
 $\overrightarrow{P}$ is therefore
$$
\widehat{\overrightarrow{P}}=
\pmatrix{0\ccr
\vec{\delta}
\ccr
-\vec{\delta}\times
\displaystyle{\vec{x}\over4\kappa}
-\displaystyle{\vec\delta\cdot\vec{J}^T\over\gamma}
+{t\over2\kappa\gamma}\vec{\delta}\times\vec{J}^T
\cr}.
\equation
$$

Some combination of conformal generators can still be  Killing.
A look on the explicit expressions (7.4)-(7.6) shows that
this happens indeed for
$$\eqalign{
\underbrace{\hbox{(hidden time translation)}}_{\widehat{\cal H}}\hfill
\cr
+\;
\big(\2{B^{ext}}\big)^2\times\underbrace{\hbox{(hidden
expansion)}}_{\widehat{\cal K}}\hfill
\cr
-\;
\big(\2{B^{ext}}\big)\times\underbrace{\hbox{(hidden rotation)}
}_{\widehat{\cal R}},\hfill
\cr}
\equation
$$
which project in fact to an {\it ordinary time translation}, $H$.
This latter lifts therefore to the metric of Case B as
$$
\hat{H}=
\pmatrix{-\epsilon
\ccr
0
\ccr
-\2\epsilon\big({\vec{J}^T\over\gamma}\big)^2
\cr}.
\equation
$$

Let us note that the formulae  referred to above are only valid in
the rest frame $\vec{J}^T=0$;
the general formulae can be obtained by a boost.

Finally, the only solutions of the Killing equation
(5.6) are combinations of these generators. The
conformal vectors of $\hat{g}_{\mu\nu}$ and of $\widetilde{g}_{\mu\nu}$
are in fact in bijection by means of the ``export/import'' map (7.1).
But on Minkowski space, the only Bargmann-conformal vectors are those
in the Schr\"odinger algebra.
Collecting our results, we state

\goodbreak
\kikezd{THEOREM2}. {\it The $\xi$-preserving isometries of the
Bargmann space of Case B form the $7$-dimensional Lie algebra,
generated by ordinary space ($\hat{P}$) and time ($\hat{H}$)
translations, vertical translations ($N$),
hidden rotations ($\widehat{\cal R}$) and hidden boosts
($\widehat{\cal G}$).
Their commutation relations read}
$$\matrix{
[\widehat{\cal G}_i,\widehat{\cal G}_j]\hfill&=&0,
\hfill
&\big[\widehat{P}_{i},\widehat{P}_{j}\big]&=&
-{1\over2\kappa}\epsilon_{ij}N,\quad\hfill
&[\widehat{P}_i,\widehat{\cal G}_{j}]\hfill&=&
\delta_{ij}{N},\hfill
\ccr
[\widehat{\cal G}_i,\widehat{\cal R}]\hfill&=&
\epsilon_{ij}\widehat{\cal G}_{j},\quad\hfill
&[\widehat{P}_i,\widehat{\cal R}]\hfill&=&
\epsilon_{ij}\widehat{P}_{j},\hfill
&&&\hfill
\ccr
[\widehat{H},\widehat{\cal R}]\hfill&=&0,\hfill
&[\widehat{H},{\cal G}_i]\hfill&=&\widehat{P}_i,\hfill
&[\widehat{H},\widehat{P}_i]\hfill&=&0.\hfill
&&&\hfill
\cr}
\equation
$$
(The vertical translation, $N$, commutes with all generators).
These
are the commutation relations of the extended Galilei group,
with the exception of that,
unlike ordinary translations, the lifted translations
 do not commute.
These latters do not form hence a subalgebra on their own
but belong rather to a three parameter subalgebra identified as
 the Heisenberg algebra,
i.e. the central extension
of space-time translations with the vertical translation.

Projecting these vector fields into ordinary space-time,
we recover the $6$ symmetries found in [6].
By (5.3), the projections satisfy the same commutation relation
 (5.11), except for that the central extension is lost under
the projection, $N\to 0$.

\kikezd{The Lifting problem}.
 Conversely, let us start with the projected vectorfields and
lift them to the Bargmann space w. r. t. the metric
$\hat{g}_{\mu\nu}$.
The lifts have non-trivial fourth components,
which come from the condition
that the transport terms ({\it alias} external fields)
be symmetric in w. r. t. the action on ordinary space-time.
 Let us illustrate this point on the example of  the ordinary
space translations.  Each of
$P_{1}=\p_{1}$ and $P_{2}=\p_{2}$ is a
symmetry~: the condition (4.2) is verified with
$$
\Upsilon_{i}
=
-{1\over2\kappa}\epsilon_{ij}\Big(\gamma x^j- tJ^T_{j}\Big)+C_{i},
\equation
$$
where $C_{i}$ is an arbitrary constant.
Each of the translations can be lifted therefore
{\it individually} to Bargmann space.
No choice of the constants $C_{i}$ allows to lift
the {\it algebra} of planar translations
{\it isomorphically},
though, since this is forbidden by
 the cohomological obstruction [23]  [24] referred
to above. Condition (4.4) would require in fact
$$
F_{\alpha\beta}^{ext}P^\alpha_{1}
P^\beta_{2}=B^{ext}\equiv{\gamma\over2\kappa}\neq0.
$$
But $\Upsilon_{0}=0$
since $P_{1}$ and $P_{2}$ commute; a contradiction.

According to (4.1) [or (4.2)], the lift of each symmetry is only unique
up to a constant.
The ambiguity can be eliminated by requiring that the
algebraic structure be (as much as possible) preserved.
For example, the Lie bracket
$$
\big[\hat{P}_{1}, {\cal R}\big]=\hat{P}_{2}
$$
fixes the constant in
$\hat{P}_{2}$, etc. Note that the constant in the lift of time
translations is {\it not} fixed as long as we only consider the
isometries, because $\hat{H}$ is not the Lie bracket
of any two isometries, cf. (5.11).
For fixing its constant, we must consider the
isometries as a subalgebra of the conformal vectors. Then the
commutation relation in (7.7) in the Appendix
do fix $\hat{H}$ uniquely, as in (5.10), upon use of (5.9).
Remarkably, the ``good lifts'' coincide with those obtained using the
``export/import'' procedure above. Fixing the lifts plays, as we
explain in the next Section, an important role in deriving the
conserved quantities.

\goodbreak
\chapter{Conserved quantities}

The lack of a variational principle forces us to use a mixed aproach,
presented in Ref. [12].
We only consider  Cases B, since Case A
has been discussed in Refs. [12].
Applying the method of Ref. [26] to the ``partial action'' $S$
(3.6), yields the symmetric energy-momentum tensor
$$
\vartheta_{\mu\nu}=2{\delta S\over\delta\hat{g}^{\mu\nu}},
\equation
$$
which also satisfies
$
\nabla_{\mu}\vartheta^{\mu\nu}+j_{\mu}f^{\mu\nu}=0.
$
Using our FCI (2.6), we see that the second term vanishes owing to
the antisymmetry.
The tensor $\vartheta^{\mu\nu}$ is hence itself conserved,
$$
\nabla_{\mu}\vartheta^{\mu\nu}=0.
\equation
$$
Our $\vartheta^{\mu\nu}$ is {\it not} traceless, though,
since the theory is {\it not} conformally symmetric.

Let $\hat{X}^\mu$ now be a Killing vector of some Brinkmann
metric $\hat{g}_{\mu\nu}$, and consider the current
$$
k^\mu=\vartheta^\mu_\nu \hat{X}^\nu.
\equation
$$
$k^\mu$ is gauge-invariant by construction.
Furthermore,
$$
\nabla_\mu(\vartheta^\mu_\nu \hat{X}^\nu)=
(\nabla_\mu\vartheta^\mu_\nu)\hat{X}^\nu+
\2\,\vartheta^{\mu\nu}\,L_{\hat{X}}g_{\mu\nu}=0
$$
since $L_Xg_{\mu\nu}=0$.
The current  $k^\mu$ is therefore conserved,
$
\nabla_\mu k^\mu=0$.
If $\hat{X}^\mu$ is also $\xi$-preserving, one can show
that the current $(k^\mu)$  projects into a three-current
$$
(K^\alpha)=(K^t,\vec{K})
\equation
$$
on ordinary space-time, $Q$.
The projected current is thus also conserved.
Hence

\kikezd{THEOREM3}. {\it For each isometry $\hat{X}$,
the quantity}\foot{The notation ${\cal{Q}}_X$ is, strictly
speaking, an abuse,
since the conserved quantity actually depends on the lift
$\hat{X}^\mu$, and not only on the space-time vector $X^\alpha$.}
$$
{\cal{Q}}_X
=
\int{
\vartheta_{\mu\nu}\hat{X}^\mu\xi^\nu\,d^2\!\vec{x}},
\equation
$$
{\it is conserved, provided all currents vanish at infinity}.

\vskip2mm
Remembering that in a local frame the lift $\hat{X}^\mu$
is decomposed as $\hat{X}^\mu=(X^\alpha, -W)=
(X^\alpha,-A_{\alpha}^{ext}X^{\alpha}+\Upsilon)$,
we get

\kikezd{COROLLARY}. {The conserved quantities admit the gauge-invariant
decomposition
$$
Q_{X}=\int\underbrace{\big[\vartheta_{\alpha s}X^\alpha
-(A^{ext}_{\alpha}X^\alpha)\vartheta_{ss}\big]}_{
\vartheta_{\mu\nu}\overline{X}^\mu\xi^\nu}d\vec{x}^2
+\int\Upsilon\vartheta_{ss}\,d\vec{x}^2,
\equation
$$
where $\Upsilon$ is the response of the symmetrical
external field to the symmetry $X$ in Eq. (4.2).
The second term  represents
here  the contribution of the symmetric external field
 to the conserved quantity,
called the ``spin from isospin'' phenomenon} [15], [19].

Varying the ``partial action'' (3.5), a rather tedious calculation
similar to that in Ref. [12]  yields the energy momentum tensor
of the lifted Manton system,
$$
\eqalign{\vartheta_{\mu\nu}
&=\smallover1/3\big(
(\D_\mu\phi)^*\,\D_\nu\phi+\D_\mu\phi\,(\D_\nu\phi)^*\big)
-
\smallover1/6\left(\phi^*\,\D_\mu \D_\nu\phi
+\phi\,(\D_\mu \D_\nu\phi)^*\right)
\cr
&+\smallover1/6|\phi|^2\left(R_{\mu\nu}-{R\over 6}\hat{g}_{\mu\nu}\right)
-
\smallover1/6\hat{g}_{\mu\nu}\,\left(\D_\sigma\phi(D^\sigma\phi)^*\right)
\cr
&-\smallover1/4\hat{g}_{\mu\nu}\left(f_{\rho\sigma}f^{\rho\sigma}\right)
-f_{\mu\rho}f^{\rho}_{\ \nu}
-\hat{g}_{\mu\nu}{\lambda\over4}\big(
-{1\over2}+{1\over3}\vert\phi\vert^2-{1\over6}\vert\phi\vert^4\big)
\ccr
&-j^T_{\mu}j^T_{\nu}+\2\hat{g}_{\mu\nu}\big(j^T_{\sigma}{j^T}^{\sigma}\big).
\cr}
\equation
$$

Since each of the currents $K^\alpha$ vanish at
infinity, setting
$\Lambda=\lambda+\big({\gamma\over\kappa}\big)^2$,
THEOREM3 yields the conserved quantities
$$\matrix{
{n}=\gamma^2\displaystyle{\int\big[1-\vert\Phi\vert^2\big]\,d^2\vec{x}}
=\gamma\displaystyle\int\big[{{\cal B}\over2\kappa}\big]\,d^2\vec{x},
\hfill
&\hbox{part. number}\hfill
\cr\ccr
{p}_{i}=
\gamma\displaystyle\int\bigg[
{\cal J}_{i}-J^T_{i}\vert\phi\vert^{2}+
\Big\{\epsilon_{ij}\big({x}^j-t{J^T_{j}\over\gamma}\big)\Big\}{\cal B}
\bigg]
d^{2}\vec{x},\hfill
&\hbox{momentum}\hfill
\cr\ccr
{h}=\displaystyle{\int\bigg[
\2\big\vert\vec{D}\phi\big\vert^{2}
-
\2\big\vert\vec{J}^T\big\vert^2\vert\phi\vert^2\hfill
+
{\Lambda\over8}\big(1-\vert\phi\vert^{2}\big)^{2}}\hfill
\cr
\qquad+\Big\{-\big(\vec{x}\times\vec{J}^T\big)\Big\}{\cal B}
\bigg]d^2x,\hfill
&\hbox{energy}\hfill
\cr\ccr
{m}=\gamma\displaystyle\int
\bigg[\vec{x}\times\big(\vec{\cal
J}-\vec{J}^T\vert\phi\vert^{2}\big)
-{t\over\gamma}\vec{J}^{T}\times\vec{\cal J}
\hfill&\hbox{hidd. ang. mom.}
\ccr
\qquad+\Big\{-\2r^2+{t\over\gamma}
(\vec{x}\cdot\vec{J}^T)
-\2\big({t\over\gamma}\big)^2\vert\vec{J}^T\vert^{2}\Big\}{\cal B}\bigg]
\,d^{2}\vec{x},\hfill
&
\cr}
\equation
$$
(The conserved quantities
associated with ``hidden boosts''
are not illuminating and are therefore not reproduced here).

These quantities, obtained  here from first principles
and without any further ``improvement''
 are identical to those found before
[6], [8].
Note that in the frame $\vec{J}^T=0$ our ${\cal M}$
becomes the ordinary angular momentum in Refs. [6] and [8].
These expressions nicely illustrate the
``spin from isospin'' phenomenon [15]~: the expressions
in the curly brackets are the $\Upsilon$s in the symmetry
definition (4.2).

The Poisson brackets of the conserved quantities (6.8)
were calculated in Ref. [6]. They
 verify the same commutation relations as the
$\xi$-preserving isometries of metric B, listed in
Eq. (5.11).
The algebraic structure of the conserved quantities reflects
hence that of Bargmann space vectors: in Souriau's terminology [23],
the ``moment map'' is equivariant for the centrally extended algebra
rather than for the projected algebra.
For the momenta in particular, the anomalous commutation relation
(1.5) is recovered.
(This latter relation can also be understood by observing
that conserved quantities associated to
``hidden translations'' and ``hidden boosts''  satisfy (5.7)).

\goodbreak
\kikezd{Acknowledgements}.
It is a pleasure to thank Professors
J.~Balog, C.~Duval and P.~Forg\'acs
for illuminating discussions. One of us (M.~H.) acknowledges
the {\it Laboratoire de Math\'ema\-thi\-ques et de Physi\-que Th\'eorique}
of Tours University for hospitality and the French Government
 for a doctoral scholarship.

\vfill\eject
\vskip4mm
\goodbreak
\centerline{\bf References}

\vskip-1mm
\reference
N.~S. Manton, 
{\it First order vortex dynamics}.
{\sl Ann. Phys}. (N. Y.). {\bf 25}, 114 (1997).
 Further references are found in Ref. [6].

\reference
R.~Jackiw and S-Y.~Pi, 
{\it Soliton solutions to the Gauged Nonlinear Schr\"odinger Equation
on the Plane}.
{\sl Phys. Rev. Lett}. {\bf 64}, 2969 (1990);
{\it Classical and quantal non-relativistic Chern-Simons theory}.
{\sl Phys. Rev}. {\bf D42}, 3500 (1990).
For reviews, see
R. Jackiw and S-Y. Pi,
{\it Self-Dual Chern-Simons Solitons}.
{\sl Prog. Theor. Phys. Suppl}. {\bf 107}, 1 (1992),
or
G. Dunne, {\it Self-Dual Chern-Simons Theories}.
Springer Lecture Notes in Physics. New Series: Monograph 36. (1995).

\reference 
S. M. Girvin,
{\it Towards a Landau-Ginsburg theory of the  FQHE}.
{\sl The Quantum Hall Effect},
edited by R. E. Prange and S.~M.~Girvin (Springer Verlag,
N. Y. 1986), Chapt. 10.
See also S.~M.~Girvin and A.-H.~MacDonald,
{\it Off-diagonal long-range order, oblique confinement, and
the fractional quantum Hall effect}.
{\sl Phys. Rev. Lett}. {\bf 58}, 303 (1987).
S. C. Zhang, T. H. Hanson and S. Kivelson,
{\it Effective-field theory model for the fractional quantum Hall
effect}. {\sl Phys. Rev. Lett}. {\bf 62}, 307 (1989).

\goodbreak
\reference 
 G.~Morandi, {\it Quantum Hall Effect. Topological problems in condensed
matter
 physics}.
Napoli~: Bibliopolis (1988);
 {\it Quantum Hall Effect}. Ed. M. Stone.
Singapore~: World Scientific (1992).

\reference 
M.~Hassa\"\i ne, P. A.~Horv\'athy et J.-C.~Yera,
{\it Vortices in the Landau--Ginzburg model of the Quantized Hall
Effect}. {\sl Journ. Phys}. {\bf A31}, 9073-79 (1998).
\goodbreak
\reference 
M.~Hassa\"\i ne, P. A.~Horv\'athy et J.-C.~Yera,
{\it Non-relativistic Maxwell-Chern-Simons Vortices}.
{\sl Ann.  Phys}. (N. Y.). {\bf 263}, 276-294 (1998).

\reference 
M.~Le Bellac and J.-M.~L\'evy-Leblond,
{\it Galilean electromagnetism}.
{\sl Il Nuovo Cimento} {\bf 14B}, 217 (1973);
 C.~R.~Hagen,
{\it A new gauge theory without an elementary photon}.
{\sl Ann. Phys}. (N. Y.). {\bf 157}, 342 (1984);
{\it Galilean-invariant gauge theory}.
{\sl Phys. Rev}. {\bf D31}, 848 (1985).

\goodbreak
\reference 
N.~S. Manton and S.~M. Nasir,
{\it Conservation laws in a first--order dynamical system of vortices}.
Cambridge preprint DAMTP-1998-80.
To appear in {\sl Nonlinearity}.

\reference 
Z.~F.~Ezawa, M.~Hotta and A.~Iwazaki,
{\it Breathing Vortex Solitons in Nonrelativistic Chern Simons Gauge
Theory}. {\sl Phys. Rev. Lett}. {\bf 67}, 411 (1991);
{\it Nonrelativistic Chern-Simons vortex solitons in external
magnetic field}. {\sl Phys. Rev}. {\bf D44}, 452 (1991).
See also R.~Jackiw and S-Y.~Pi,
{\it Semiclassical Landau Levels of Anyons}.
{\sl Phys. Rev. Lett}. {\bf 67}, 415 (1991);
{\it Time-dependent Chern-Simons solitons and their quantization}.
{\sl Phys. Rev}. {\bf D44}, 2524 (1991).
 M.~Hotta,
{\it Imported Symmetry and Two Breathing Modes in Chern-Simons
Theory with External Magnetic Field}.
 {\sl Prog. Theor. Phys}. {\bf 86}, 1289 (1991).

\goodbreak
\reference 
I.~Barashenkov and A.~Harin,
{\it Non-relativistic Chern-Simons theory for the repulsive Bose gas}.
 {\sl Phys. Rev. Lett}. {\bf 72}, 1575 (1994).

\reference 
C.~Duval, G.~Burdet, H-P.~K\"unzle and M.~Perrin,
{\it Bargmann structures and Newton-Cartan theory}.
{\sl Phys. Rev}. {\bf D31}, 1841 (1985);
C.~Duval, G.~Gibbons and P.~Horv\'athy,
{\it Celestial mechanics, conformal structures, and gravitational
waves}.
{\sl Phys. Rev}. {\bf D43}, 3907 (1991).

\reference  
C.~Duval, P.~A.~Horv\'athy and L.~Palla,
{\it Conformal symmetry of the coupled Chern-Simons and gauged
nonlinear Schr\"odinger equations}.
{\sl Phys. Lett}. {\bf B325}, 39 (1994).

\reference 
C.~Duval, P. A. Horv\'athy and L. Palla,
{\it Conformal properties of Chern-Simons vortices in external fields}.
{\sl Phys. Rev}. {\bf D50}, 6658 (1995).

\reference 
R.~Jackiw, {\it Scaling symmetries}
{\sl Physics Today} (1980);
U.~Niederer, {\it The maximal kinematical invariance group of the free
Schr\"odinger equation}.
{\sl Helvetica Physica Acta} {\bf 45}, 802 (1972);
C.~R.~Hagen, {\it Scale and conformal transformations in
Galilean-covariant field theory}.
{\sl Phys. Rev}.  {\bf D5}, 377 (1972).

\reference 
P. Forg\'acs and N. Manton,
{\it Space-time symmetries in gauge theories}.
{\sl Commun. Math. Phys.} {\bf 72}, 15 (1980);
R. Jackiw and N. Manton, 
{\it Symmetries and conservation laws in gauge theories}.
{\sl Ann. Phys}. (N. Y.) {\bf 127}, 257 (1980).

\reference 
T.~Ka\l u{z}a, {\sl Sitzungsber. Preuss. Acad. Wiss. Phys. Math}.
{\bf K1}, 996 (1916);
O.~Klein, {\it }
{\sl Z. Phys}. {\bf 37}, 895 (1926).
For a modern account, see, e. g.,
R.~Coquereaux and A.~Jadczyk,
{\it Riemannian geometry, fiber bundles, Kaluza-Klein Theories
and all that}. World Scientific Lecture Notes {\bf 16}.
Singapore~: World Scientific (1988).

\reference 
S.~Kobayashi and K.~Nomizu,
{\sl Foundations of differential geometry}.
Vol. {\rm I} Interscience: New York (1963).

\reference 
J.~Harnad, S.~Shnider and L.~Vinet,
{\it Group action on principal bundles and invariance conditions for
gauge fields}.
{\sl J. Math. Phys}. {\bf 21}, 2719 (1980).

\reference 
 C.~Duval and P.~A.~Horv\'athy,
 {\it Particles with internal structure: the geometry of classical
 motions and conservation laws}.
 {\sl Ann. Phys}. (N. Y.) {\bf 142}, 10-33 (1982).

\reference
H.W.~Brinkmann, 
{\it Einstein spaces which are mapped conformally on each other}.
{\sl Math. Ann}. {\bf 94}, 119 (1925).
Such metrics also provide exact string vacua, see
C.~R.~Nappi and E.~Witten,
{\it Wess-Zumino model based on a nonsemisimple group}.
{\sl Phys. Rev. Lett}. {\bf 71}, 3751 (1993);
C.~Duval, Z.~Horv\'ath and P.~A. Horv\'athy,
{\it Strings in plane-fronted graviational waves}.
{\sl Mod. Phys. Lett}. {\bf A8}, 3749-3756 (1993);

\reference 
S.~M.~Carroll, G.~B.~Field, and R.~Jackiw,
{\it Limits on a Lorentz and parity violating modification
of electrodynamics}.
{\sl Phys. Rev}. {\bf D41}, 1231 (1990).

\reference 
S.~K.~Paul and A.~Khare,
{\it Charged vortices in the abelian Higgs model with Chern-Simons term}.
{\sl Phys. Lett}. {\bf B174}, 420 (1986).

\reference 
J.~M.~Souriau, {\it Structure des syst\`emes dynamiques}.
Dunod: Paris (1970).

\reference 
D.~J.~Simms,
{\it Projective representations, symplectic manifolds
and extensions of Lie algebras}.
Marseille Lecture Notes CPT-69/P.300. (1969).

\reference 
U.~Niederer,
{\it The Maximal Kinematical Invariance Group of the Harmonic
Oscillator}.
{\sl Helvetica Physica Acta} {\bf 46}, 191 (1973).

\reference  
J.-M.~Souriau,
{\it Mod\`ele de particule \`a spin dans le champ \'electromagn\'etique
et gravitationnel}.
{\sl Ann. Inst. H.~Poincar\'e} {\bf 20A}, 315 (1974).

\vfill\eject
\chapter{Appendix}

The Bargmann-conformal transformation $\Psi(t,\vec{x},s)\to (T,\vec{X},S)$
which takes  Metric B into that
of Minkowski space reads explicitly
$$\eqalign{
&T=\smallover{2}/{B^{ext}}\tan\smallover{2}/{\gamma B^{ext}}\,t,\hfill
\ccr
&
X_{k}=x_{k}
-\epsilon_{kl}{{E^{ext}}_l\over{B^{ext}}}t
-\epsilon_{kl}\left(x_{l}-\epsilon_{lm}
{{E^{ext}}_m\over {B^{ext}}}t\right)\tan\smallover{2}/{\gamma
B^{ext}}\,t,\hfill
\ccr
&S=s
+\epsilon_{lm}x_{l}
{{E^{ext}}_m\over {B^{ext}}}
-{1\over\kappa}\big({\vec{E}^{ext}\over{B^{ext}}}\big)^2t
+\smallover1/{2\gamma}t\,\vec{x}\cdot\vec{E}^{ext}
\hfill
\ccr
&\hskip20mm
-\smallover1/{2\gamma}{B^{ext}}\big(x_{l}-\epsilon_{lm}
{{E^{ext}}_m\over{B^{ext}}}t\big)^2
\tan\smallover{2}/{\gamma B^{ext}}\,t\hfill
\cr}
\equation
$$

The ``hidden Schr\"odinger algebra'' is obtained by ``importing''
the Schr\"odinger algebra (5.4) by (7.1).
The isometries act on Bargmann space as
\goodbreak
$$\matrix{
\widehat{\overrightarrow{\cal{P}}}=\quad\cos{1\over4\kappa}\,t
\pmatrix{0
\ccr
\cos{1\over4\kappa} t\,\Gamma_{1}+\sin{1\over4\kappa} t\,\Gamma_{2}
\ccr
-\sin{1\over4\kappa} t\,\Gamma_{1}+\cos{1\over4\kappa} t\,\Gamma_{2}
\ccr
f
\cr}\qquad
&\hbox{``hidden translations''},\hfill
\ccr
\widehat{\overrightarrow{\cal{G}}}={4\kappa\over\gamma}\sin{1\over4\kappa} t
\pmatrix{0
\ccr
\cos{1\over4\kappa} t\,\beta_1+\sin{1\over4\kappa} t\,\beta_{2}
\ccr
-\sin{1\over4\kappa} t\,\beta_{1}+\cos{1\over4\kappa} t\,\beta_{2}
\ccr
g
\cr}\qquad
&\hbox{``hidden boosts''},\hfill
\cr\ccr
\quad\qquad\qquad
\widehat{\cal R}=\pmatrix{0
\ccr
\Omega\big(-x_2+J_2^T{t\over\gamma}\big)
\ccr
\Omega\big(x_1-J_1^T{t\over\gamma}\big)
\ccr
h
\cr}\;
\hfill
&\hbox{``hidden rotations''},\hfill
\cr\ccr
N=\pmatrix{0
\ccr
0
\ccr
0
\ccr
\eta
\cr}\qquad
&\hbox{``vertical translations''},\hfill
\cr
}
\equation
$$
where
$\vec{\Gamma},\, \vec{\beta}\in\IR^{2}$,
$\Omega,\, \eta\in\IR$.
In these formul{\ae},  $f$, $g$ and $h$ are shorthands
for the complicated expressions
$$\matrix{
f=&{1\over4\kappa\cos{1\over4\kappa} t}&\left[
-\sin^2\big({t\over4\kappa}\big)\vec\gamma\times\vec{x}+
+\sin\big({t\over4\kappa}\big)\cos\big({t\over4\kappa}\big)\,
\vec{x}\cdot\vec{\Gamma}
\right.\hfill
\ccr
&&\left.+{t\over\gamma}\,\vec{\Gamma}\times\vec{J}^T
-{4\kappa\over\gamma}\,\cos^2\big({t\over4\kappa}\big)
\vec\gamma\cdot\vec{J}^T
+{4\kappa\over\gamma}
\sin\big({t\over4\kappa}\big)\cos\big({t\over4\kappa}\big)\,
\vec{\Gamma}\times\vec{J}^T\right],\hfill
\cr\cr
g=&{1\over4\kappa\sin{1\over4\kappa} t}&\left[
-\cos^{2}\big({t\over4\kappa}\big)\,\vec{x}\cdot\vec{\beta}
+\sin\big({t\over4\kappa}\big)\cos\big({t\over4\kappa}\big)\,
\vec\beta\times\vec{x}
\right.\hfill
\ccr
&&\left.
+{t\over\gamma}\,\vec{\beta}\cdot\vec{J}^T
+{4\kappa\over\gamma}\,\sin^2\big({t\over4\kappa}\big)\vec\beta\times\vec{J}^T
-{4\kappa\over\gamma}
\sin\big({t\over4\kappa}\big)\cos\big({t\over4\kappa}\big)\,
\vec{\beta}\cdot\vec{J}^T\right],\hfill
\cr\cr
h=&\Omega&\left[
-{t\over4\kappa\gamma}\big(\vec{x}\cdot\vec{J}^T\big)
+{1\over4\kappa}\big({t\over\gamma}\big)^{2}(\vec{J}^T)^{2}
-{\vec{x}\times\vec{J}^T\over\gamma}\right].\hfill
\ccr}
\equation
$$

 ``Hidden dilatations'' and ``hidden expansions''
and (somewhat surprisingly) ``hidden time translations''
are conformal but not Killing.
Setting
$\tau={t\over 4\kappa}$, they read\foot{For simplicity, we only
present the formul{\ae} valid in the rest frame $\vec{J}^{T}=0$.
The general expressions (which would take several pages to write)
can be found by boosting those in (7.4).}
$$\matrix{
\widehat{\cal H}=\hfill
&\hbox{hidden time translation}=\hfill
\ccr
&\pmatrix{
-\gamma\cos^2{\tau}
\ccr
{\gamma\over4\kappa}\cos{\tau}\big(x_{1}\sin{\tau}
-x_{2}\cos{\tau}\big)
\ccr
{\gamma\over4\kappa}\cos{\tau}\big(x_{1}\cos{\tau}
+x_{2}\sin{\tau}\big)
\ccr
-{r^2\gamma\over32\kappa^2}
\cos2{\tau}
\cr}.
\cr}
\equation
$$
$$\matrix{
\widehat{\cal K}=\hfill
&\hbox{hidden expansion}=\hfill
\ccr
&\pmatrix{
-({16\kappa^{2}\over\gamma})\sin^2{\tau}
\ccr
-{4\kappa\over\gamma}\sin{\tau}\big(
x_{1}\cos{\tau}
+x_{2}\sin{\tau}\big)
\hfill
\ccr
{4\kappa\over\gamma}\sin{\tau}\big(
x_{1}\sin{\tau}
-x_{2}\cos{\tau}\big)
\hfill
\ccr
{r^2\over2\gamma}
\cos2{\tau}
\cr}.\hfill
\ccr}
\equation
$$
$$
\matrix{
\widehat{\cal D}=
\hfill
&\hbox{hidden dilatation}=\hfill
\cr\cr
&-{1\over2}\pmatrix{
4\kappa\sin2{\tau}
\ccr
x_{1}\cos2{\tau}-x_{2}\sin2{\tau}\hfill
\ccr
x_{1}\sin2{\tau}+x_{2}\cos2{\tau}
\hfill
\ccr
{r^2\over4\kappa}\sin2{\tau}
\cr}.
\hfill
\cr
}
\equation
$$

The commutation relations of the ``hidden'' quantities are those of the
Schr\"odinger algebra.
\end
$$\matrix{
[{\cal G}_i,{\cal G}_j]\hfill&=&0,\hfill
&[{\cal P}_i,{\cal P}_j]\hfill&=&0,\hfill
&[{\cal P}_i,{\cal G}_{j}]\hfill&=&{\delta_{ij}\over\gamma}{N},\hfill
\ccr
[{\cal G}_i,{\cal R}]\hfill&=&\epsilon_{ij}{\cal G}_{j},\qquad\hfill
&[{\cal P}_i,{\cal R}]\hfill&=&\epsilon_{ij}{\cal P}_{j},\qquad\hfill
&&&
\ccr
[{\cal H},{\cal G}_i]\hfill&=&{\cal P}_i,\hfill
&[{\cal H},{\cal P}_i]\hfill&=&0,\hfill
&[{\cal H},{\cal R}]\hfill&=&0,\hfill
\ccr
[{\cal H}, {\cal D}]\hfill&=&2{\cal H},\hfill
&[{\cal H},{\cal K}]\hfill&=&{\cal D},\hfill
&[{\cal D},{\cal K}]\hfill&=&2{\cal K},\hfill&
\ccr
[{\cal R},{\cal D}]\hfill&=&0,\hfill
&[{\cal R},{\cal K}]\hfill&=&0,\hfill
&[{\cal D},{\cal G}_i]\hfill&=&{\cal G}_i,\hfill
\ccr
[{\cal D},{\cal P}_i]\hfill&=&-{\cal P}_i,\hfill
&[{\cal K},{\cal G}_i]\hfill&=&0,\hfill
&[{\cal K},{\cal P}_i]\hfill&=&{\cal G}_i.\hfill
\cr
}
\equation
$$

\end